# Non-Explosive Hydrogen Burning: Where Do We Stand?

M. Arnould[1], N. Mowlavi[1] and A. Champagne[2]

[1]Institut d'Astronomie et d'Astrophysique, Université Libre de Bruxelles
C.P. 226, Bd. du Triomphe, B-1050 Bruxelles, Belgium
e-mail: *marnould@astro.ulb.ac.be* and *nmowlavi@astro.ulb.ac.be*
[2]Department of Physics and Astronomy, University of North Carolina
Chapel Hill, N.C. 27599-3255, USA
and Triangle Universities Nuclear Laboratory
Duke University, Durham, N.C. 27706, USA
e-mail: *aec@tunl.tunl.duke.edu*

**Abstract:** The impact of nuclear physics uncertaintites on the abundances predicted to emerge from the cold CNO, NeNa and MgAl modes of hydrostatic hydrogen burning is discussed in the framework of a simple parametric model. In addition of being able to mimic qualitatively detailed stellar model predictions, these parametric calculations have the virtue of isolating in a crystal-clear way abundance uncertainties of purely nuclear physics origin.

# 1 Introduction

The life of a star is made of a succession of "controlled" thermonuclear burning stages interspersed with phases of gravitational contraction. The latter stages are responsible for a temperature increase, while the former ones produce nuclear energy and lead to composition changes.

The first major nuclear burning phase is the combustion of hydrogen in the central stellar regions, settling the object on the main sequence of the Hertzsprung-Russell diagram. After its exhaustion in the stellar core, hydrogen burns in peripheral layers, and the star is pushed to the Red Giant evolutionary stage (at least in absence of intense stellar winds).

Apart from its energetic importance, H burning deeply modifies the composition of the stellar interiors. The ashes of these transformations can be brought to the surface layers of the stars as a result of dredge-up episodes that can develop at specific (Red Giant and Asymptotic Red Giant) phases of the evolution of the stars, particularly in the low- to intermediate-mass range. The confrontation between observed and calculated abundances can thus provide essential clues on the stellar structure from the main sequence to the Red Giant phase, at least if the predictions are freed from nuclear physics uncertainties to the largest possible extent.

Much theoretical and laboratory effort has been put recently into improving our knowledge of many of the reactions involved in the non-explosive H-burning modes. These are (e.g. Rolfs and Rodney 1988) the pp-chains, the "cold" CNO cycles, and the NeNa and MgAl chains, the first two modes being essential energy producers, all four being of importance as far as nucleosynthesis is concerned.

In spite of this work, important uncertainties remain. This relates directly to the enormous problems the experiments have to face in this field, especially because the astrophysically relevant energies are much lower than the Coulomb barrier energies. As a consequence, the corresponding cross sections can dive into the nanobarn to picobarn abyss. In general, it has not been possible yet to measure directly such small cross sections. Theoreticians are thus requested to supply reliable extrapolations from the lowest energies attained experimentally to those of astrophysical interest.

The aim of this brief review is to evaluate the impact on various abundance predictions of the nuclear physics uncertainties still affecting the rates of some reactions involved in the non-explosive CNO, NeNa and MgAl modes. The yields are calculated by combining *in all possible ways* the lower and upper limits of all the rates for which such an information is provided. One set of calculation is also performed with "recommended rates". Note that the nuclear physics aspects of the pp-chains have been discussed at length in many recent papers in relation with the solar neutrino problem (e.g. Dzitko et al. 1995, and references therein), and will not be dealt with again here.

Our extensive abundance uncertainty analysis is performed in the framework of a parametric model assuming that H burning takes place at a constant density $\rho = 100$ gcm$^{-3}$ and at constant temperatures between $T_6 = 10$ and 80 ($T_6$ is the temperature in units of $10^6$ K). This range encompasses typical H-burning temperatures in a large variety of realistic stellar models. Initial abundances are assumed to be solar (Anders and Grevesse 1989), and the H-burning nucleosynthesis is followed until the hydrogen mass fraction $X(\mathrm{H})$ drops to $10^{-5}$.

In spite of its highly simplistic aspect, this analysis provides results that are of reasonable qualitative value, as testified by their confrontation with detailed stellar model predictions. In addition, these parametric calculations have the virtue of isolating in a crystal-clear way the abundance uncertainties that are of purely nuclear physics origin.

## 2 The CNO Cycles

The reactions involved in the CNO cycles are presented in Fig. 1. As is well known, their net result is the production of $^4$He from H, and a substantial transformation of the various C, N and O isotopes into $^{14}$N as a result of the slowness of $^{14}\mathrm{N}(\mathrm{p},\gamma)^{15}\mathrm{O}$ relative to the other reactions involved in the CNO cycles. This $^{14}$N accumulation is clearly seen in Fig. 2.

As shown in Fig. 1, three nuclides are important CNO cycle branching points. The first one is $^{15}$N. At $T_6 = 25$, $^{15}\mathrm{N}(\mathrm{p},\alpha)^{12}\mathrm{C}$ is 1000 times faster than $^{15}\mathrm{N}(\mathrm{p},\gamma)^{16}\mathrm{O}$, and the CN cycle reaches equilibrium already before $10^{-3}$ of the protons have been burned.

The second branching nuclide is $^{17}$O. The competing $^{17}\mathrm{O}(\mathrm{p},\alpha)^{14}\mathrm{N}$ and $^{17}\mathrm{O}(\mathrm{p},\gamma)^{18}\mathrm{F}$ proton capture reactions determine the relative importance of cycle II over cycles III and IV. They have been subjected recently to much experimental work (Landré et al. 1989, Berheide et al. 1993, Blackmon et al. 1995), the results of which are displayed in Fig. 3. The rates provided by Berheide et al. (1993) are upper limits only. Quite embarrassingly, they are found to be lower than the lower limits of Landré et al. (1989). The rates proposed by Blackmon et al. (1995) have much reduced error bars, and lie close to the lower limits of the rates deduced by Landré et al. (1989).

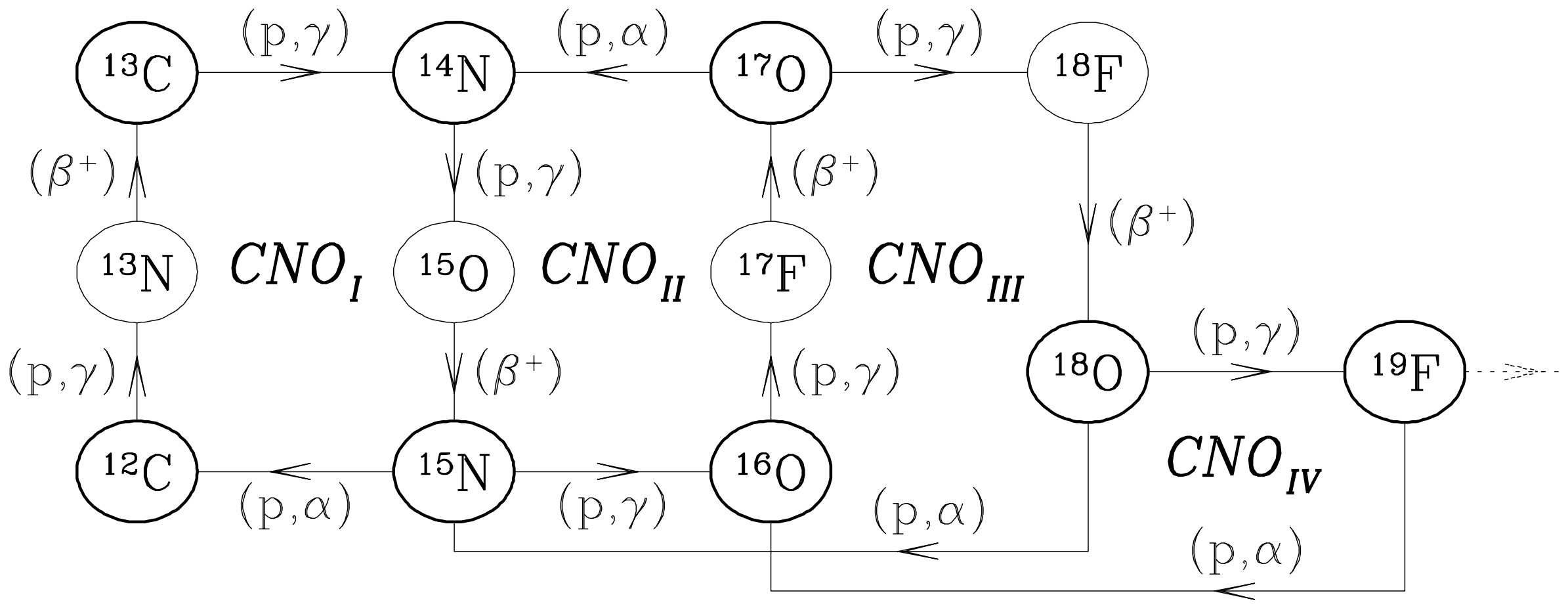

Figure 1: Reactions of the CNO cycles. Stable nuclides are enclosed in thick circles. The dashed line represents the possible leakage out of the cycles.

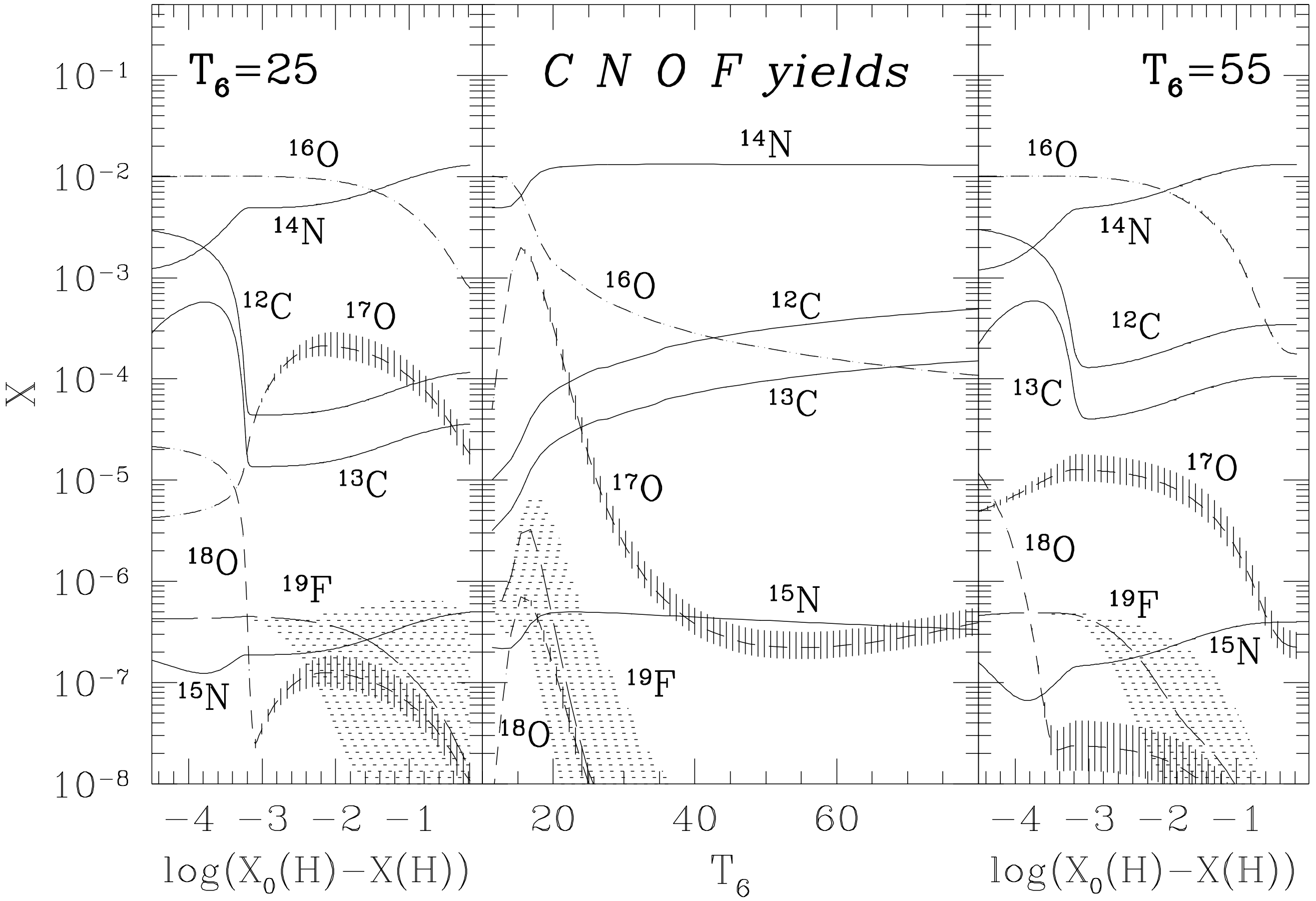

Figure 2: *Left and right panels:* Time evolution of the mass fractions of the stable nuclides involved in the CNO cycles versus the amount of hydrogen burned at constant temperatures $T_6 = 25$ and 55 and density $\rho = 100$ g/cm$^3$; *Middle panel:* Mass fractions of these nuclides at H exhaustion [X(H)=$10^{-5}$] versus $T_6$.

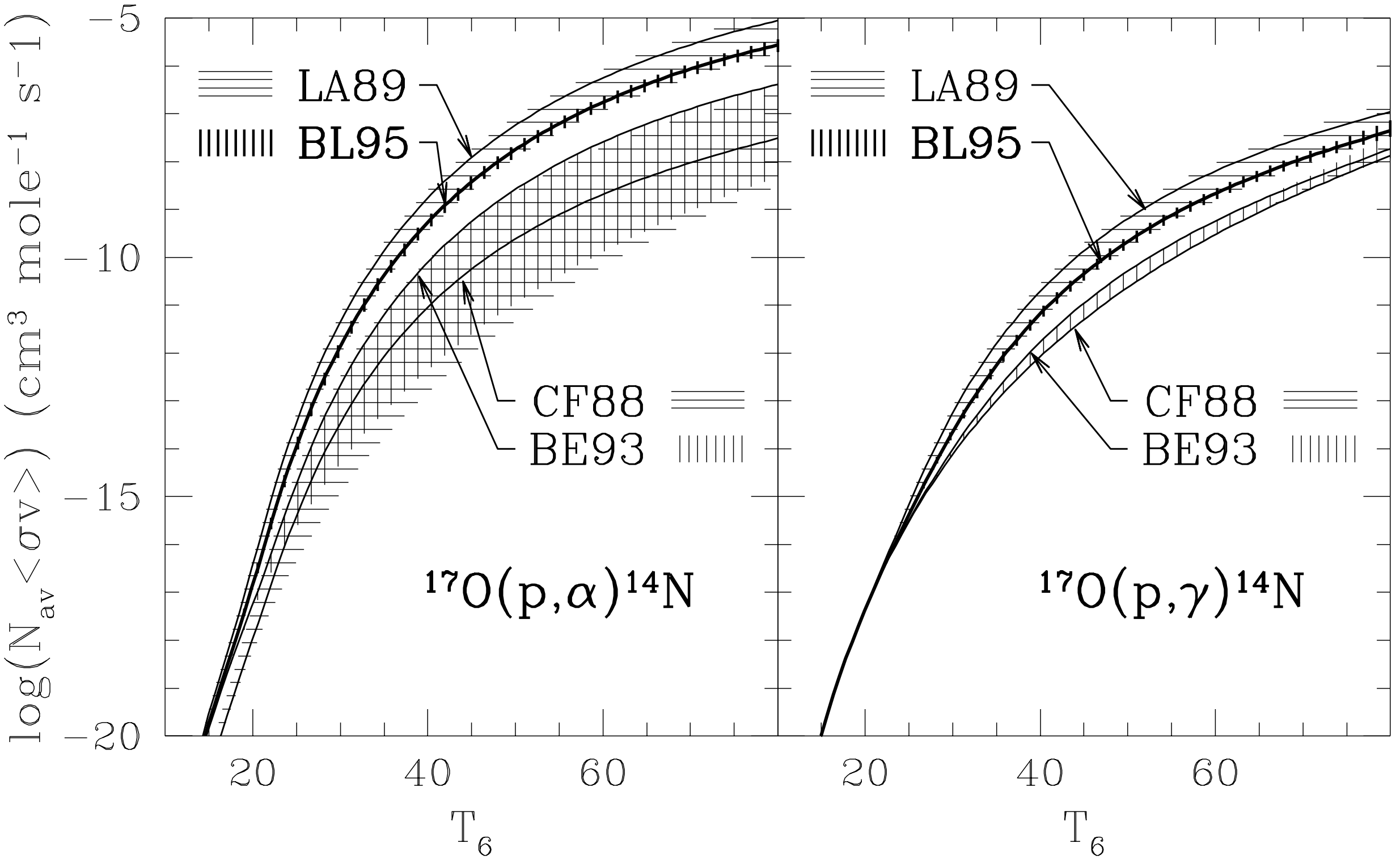

Figure 3: Temperature dependence of the rates of $^{17}O\,(p\,,\alpha)\,^{14}N$ and $^{17}O\,(p\,,\gamma)\,^{18}F$. LA89, BE93 and BL95 refer to Landré et al. (1989), Berheide et al. (1993), and Blackmon et al. (1995). Solid lines indicate recommended rates except in the BE93 case, where the line corresponds to an upper limit. Rate uncertainties are represented by hatched areas extending between the lower and upper rate limits.

The impact on the $^{17}O$ and $^{18}O$ yields of the uncertainties reported by Blackmon et al. (1995) are shown in Fig. 2. It appears that (i) the spread in the oxygen isotopic yields is now reduced to less than 7% in logarithmic units, and (ii) the oxygen isotopic composition depends drastically on the burning temperature. In particular, $^{17}O$ is produced at $T_6 \lesssim 30$, but is destroyed at higher temperatures. This has the important consequence that the amount of $^{17}O$ emerging from the CNO cycles and eventually dredged-up to the stellar surface is a steep function of the stellar mass. This conclusion could get some support from the observation of a large spread in the oxygen isotopic ratios at the surface of red giant stars of somewhat different masses (Lambert et al. 1986).

Finally, the leakage from cycle III to cycle IV is determined by the ratio of the $^{18}O\,(p\,,\gamma)\,^{19}F$ and $^{18}O\,(p\,,\alpha)\,^{15}N$ rates. At the temperatures of relevance, $^{18}O\,(p\,,\gamma)\,^{19}F$ is roughly 1000 times slower than $^{18}O\,(p\,,\alpha)\,^{15}N$ according to the compilation of Cauglan and Fowler (1988; CF88), undermining the path to $^{19}F$.

A recent analysis of the $^{19}F$ proton capture rates by Kious (1990) indicates that $^{19}F\,(p\,,\alpha)\,^{16}O$ might be slower than the $^{18}O\,(p\,,\gamma)\,^{19}F$ production channel in a range of temperatures the extent of which cannot be very precisely established, especially in vue of the uncertainties still remaining in $^{19}F\,(p\,,\alpha)\,^{16}O$ (Fig. 4). As a result, some accumulation of $^{19}F$ might be obtained, which is just impossible with the CF88 rates. Figure 2 indeed confirms that fluorine could be overproduced (with respect to solar) by up to a factor of 10 at H exhaustion if $T_6 \simeq 17$.

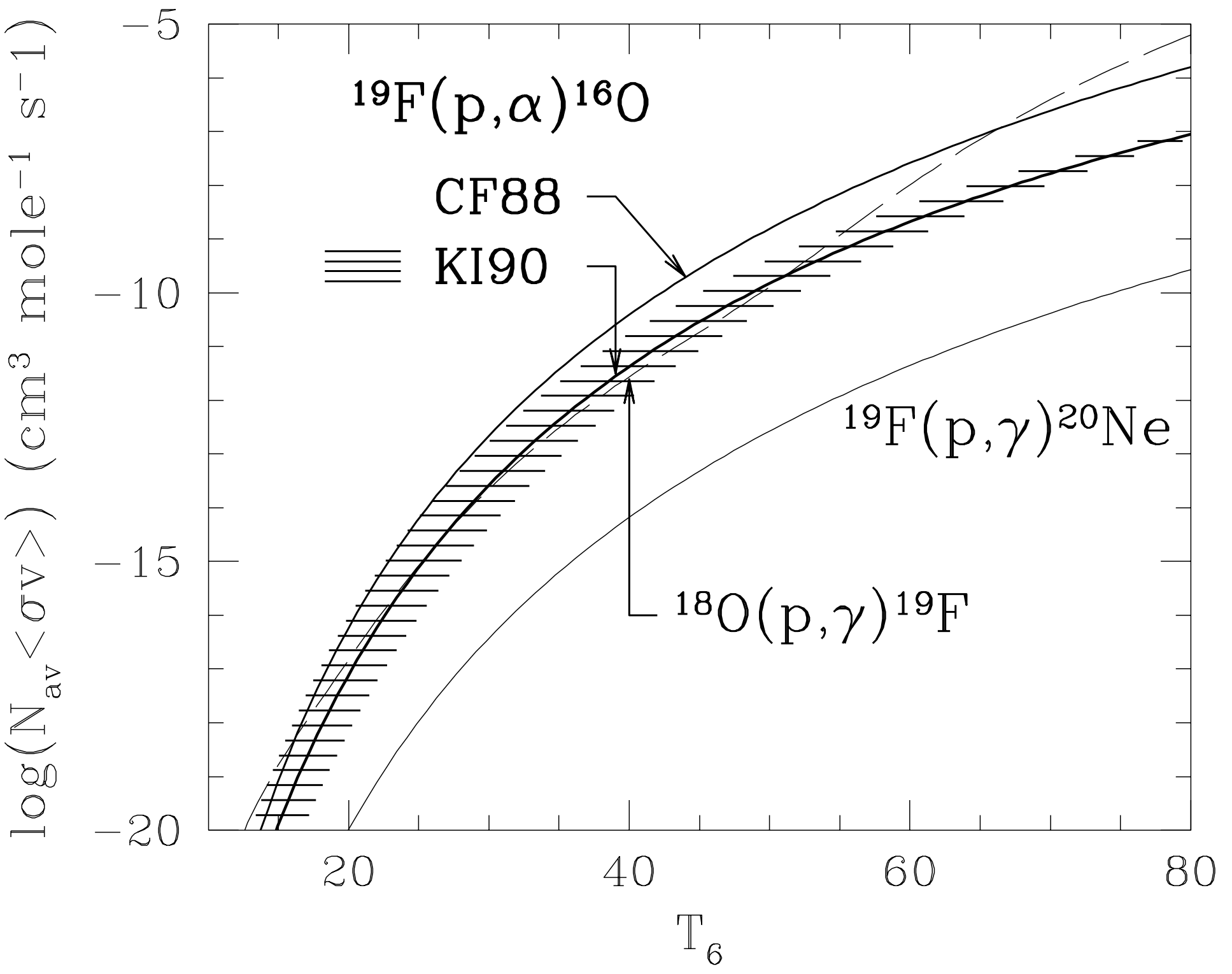

Figure 4: Same as Fig. 3, but for the reactions involved in the production and destruction of $^{19}$F by the CNO cycles. KI90 refers to Kious (1990). The rates of $^{18}O\,(p\,,\gamma)\,^{19}F$ and $^{19}F\,(p\,,\gamma)\,^{20}Ne$ are taken from CF88.

However, Fig. 2 also stresses that the maximum $^{19}$F yields that can be attained remains very poorly predictable as a direct result of the uncertainties remaining in the $^{19}F\,(p\,,\alpha)\,^{16}O$ rate. In fact, some hint that indeed fluorine can emerge in non-negligible amounts from the CNO cycles might come from the observation that slightly larger than solar fluorine abundances are present at the surface of Red Giant stars (Jorissen et al. 1992). These could possibly result from the first dredge-up mechanism (Mowlavi et al. 1995).

Finally, let us note that $^{19}F\,(p\,,\alpha)\,^{16}O$ is always much faster than $^{19}F\,(p\,,\gamma)\,^{20}Ne$. Any important leakage out of the CNO cycles to $^{20}$Ne is thus prevented, this conclusion being independent of the remaining rate uncertainties.

# 3 The NeNa Chain

The NeNa chain is illustrated in Fig. 5. Since the CF88 compilation, little in the way of new experimental information has been reported for the relevant reactions, even if the rates of several of them still exhibit large uncertainties. These result primarily from limits on the strengths of presumed, but undetected, low-energy resonances. Recently, El Eid and Champagne (1995) have presented a new set of rates which use improved estimates for these possible resonances.

These revised rates have been used in order to compute the abundances displayed in Fig. 6. A slight alteration of the initial $^{20}$Ne abundance is visible only for $T_6 \gtrsim 60$. However, an unnoticeable $^{20}$Ne destruction is sufficient to lead to a significant increase of the abundance of the rare $^{21}$Ne isotope through $^{20}Ne\,(p\,,\gamma)\,^{21}Na\,(\beta^+)\,^{21}Ne$ at $T_6 \lesssim 30$. At higher temperatures,

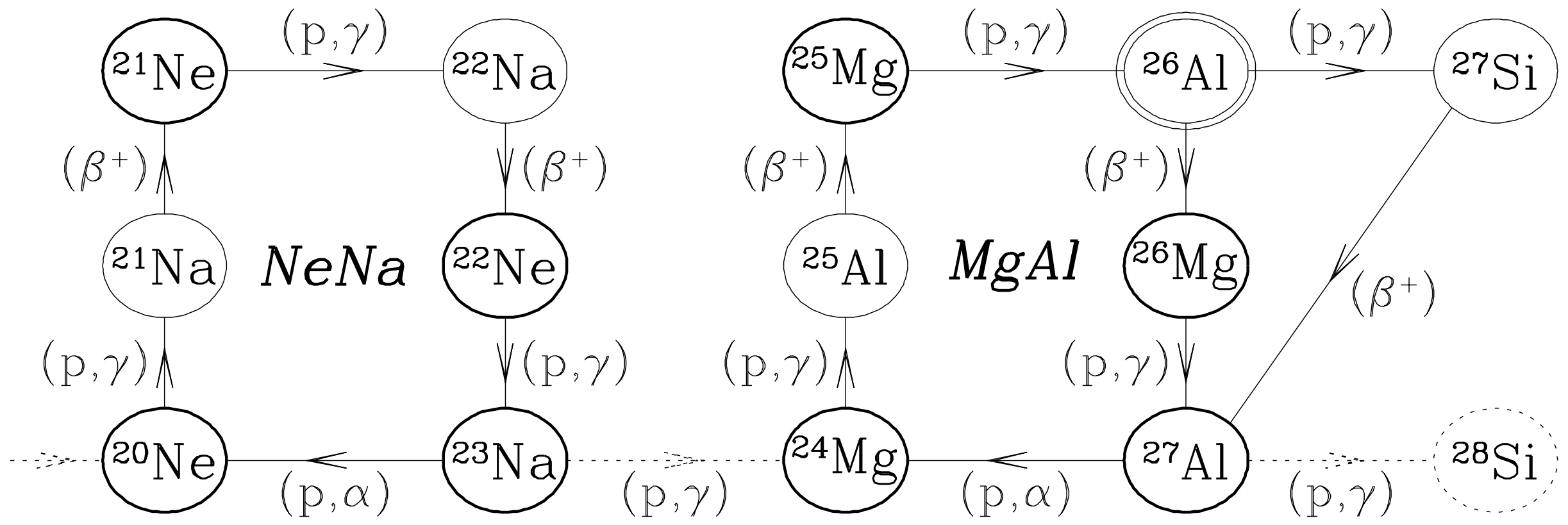

Figure 5: Same as Fig. 1, but for the NeNa and MgAl chains. The ground and isomeric states of $^{26}$Al are considered as two separate species.

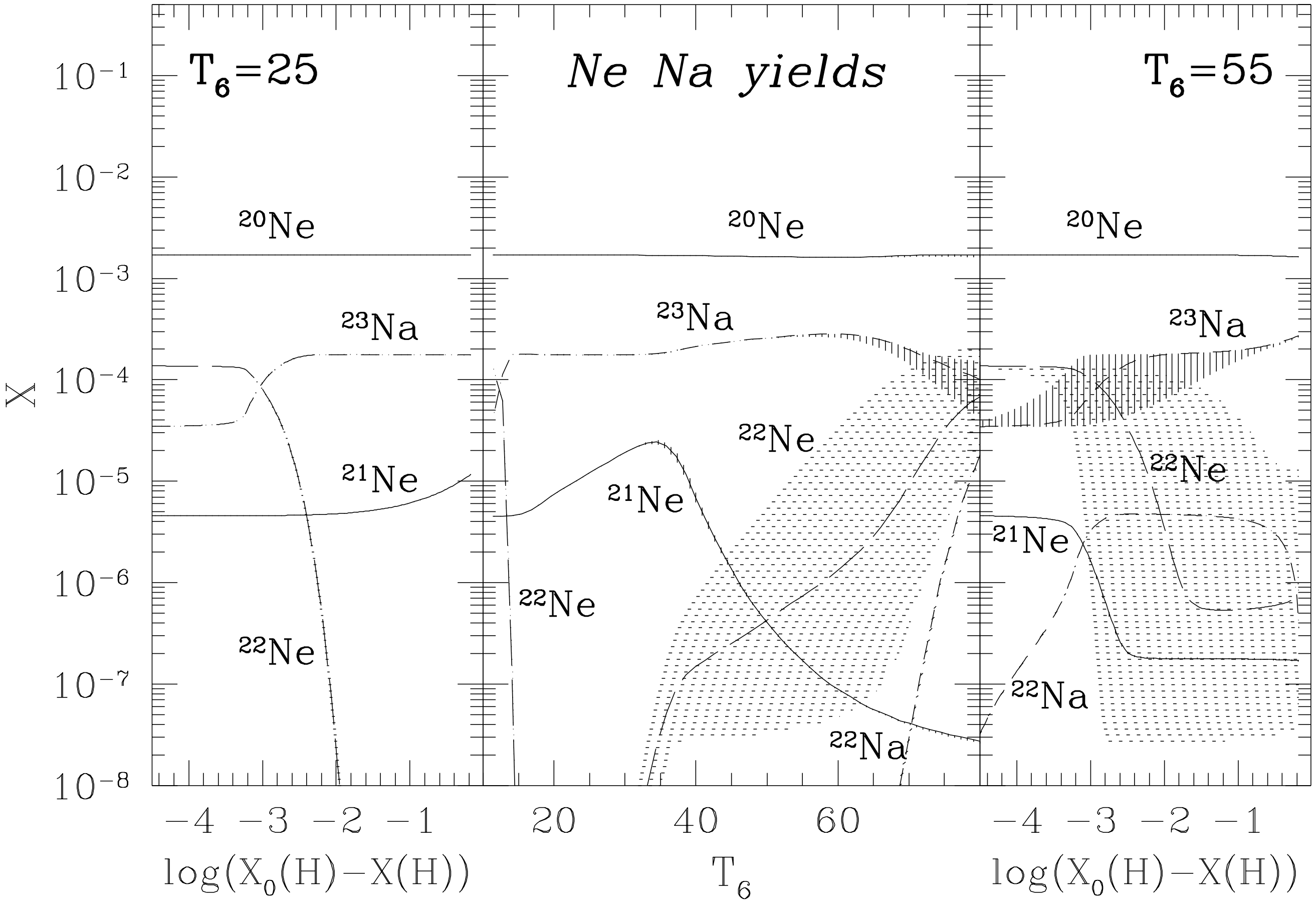

Figure 6: Same as Fig. 2, but for the nuclides involved in the NeNa chain.

$^{21}$Ne is destroyed by $^{21}\mathrm{Ne}\,(\mathrm{p}\,,\gamma)\,^{22}\mathrm{Na}\,(\beta^+)\,^{22}\mathrm{Ne}$. As a result, the $^{21}$Ne abundance at H exhaustion is maximum when the burning proceeds at a temperature in the approximate $30 \lesssim T_6 \lesssim 35$ range. This conclusion is not affected by the less than 40% uncertainty reported by El Eid and Champagne (1995) for $^{21}\mathrm{Ne}\,(\mathrm{p}\,,\gamma)\,^{22}\mathrm{Na}$, and would even remain basically unaltered with the

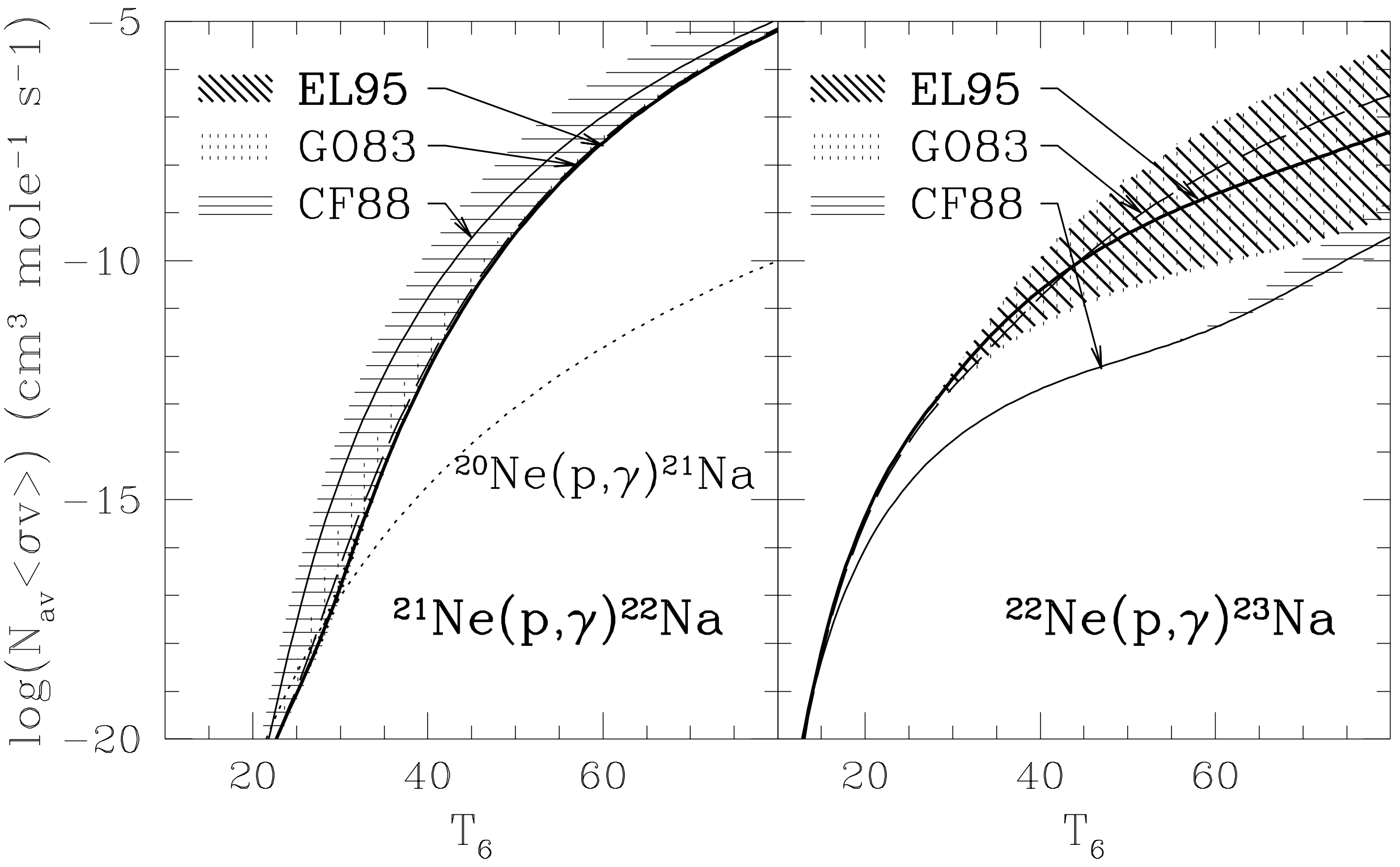

Figure 7: Same as Fig. 3, but for $^{21}$Ne(p, $\gamma$)$^{22}$Na and $^{22}$Ne(p, $\gamma$)$^{23}$Na. EL95 refers to El Eid and Champagne (1995), and GO83 to Görres et al. (1983; dashed line). The rate of $^{20}$Ne(p, $\gamma$)$^{21}$Na shown in dotted line on the left panel is taken from CF88.

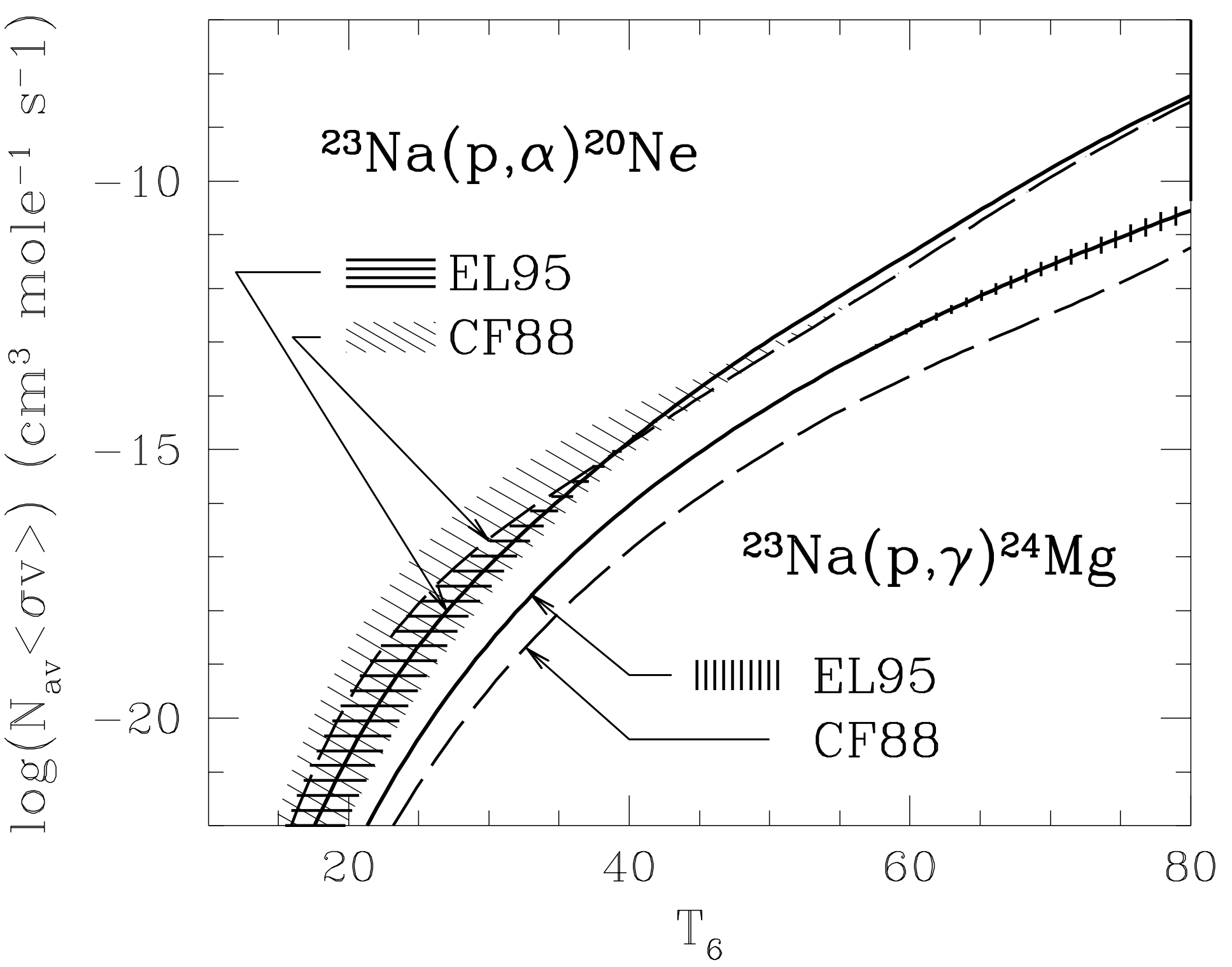

Figure 8: Same as Fig. 3, but for $^{23}$Na(p, $\gamma$)$^{24}$Mg and $^{23}$Na(p, $\alpha$)$^{20}$Ne. EL95 refers to El Eid and Champagne (1995).

adoption of the factor of up to 50 uncertainty claimed by Görres et al. (1983) (Fig. 7).

The $^{23}$Na yields from the NeNa chain has raised much interest recently following the discovery of moderate sodium overabundances at the surface of some Red Giant stars (Takeda and Takada-Hidai 1994), interpreted as the result of the dredge-up to the stellar surface of ashes of the NeNa chain. The production of $^{23}$Na results from $^{22}\mathrm{Ne}(\mathrm{p},\gamma)^{23}\mathrm{Na}$, while it can be destroyed at $T_6 \gtrsim 60$ by $^{23}\mathrm{Na}(\mathrm{p},\gamma)^{24}\mathrm{Mg}$ and by $^{23}\mathrm{Na}(\mathrm{p},\alpha)^{20}\mathrm{Ne}$. The large uncertainties remaining in the $^{22}\mathrm{Ne}(\mathrm{p},\gamma)^{23}\mathrm{Na}$ rate at $T_6 \gtrsim 50$ (Fig. 7) and, to a lesser extent, in the $^{23}\mathrm{Na}(\mathrm{p},\gamma)^{24}\mathrm{Mg}$ rate (Fig. 8) fortunately do not translate into as large uncertainties in the $^{23}$Na yields at hydrogen exhaustion. It has also to be noted that the $^{23}$Na abundance predictions of Fig. 6 based on the rates of El Eid and Champagne (1995) differ from the yields obtained with the CF88 rates (Fig. 6 in Arnould and Mowlavi 1993) by a factor of up to 3 above $40 \times 10^6$ K.

On the other hand, the relative rates of $^{23}\mathrm{Na}(\mathrm{p},\alpha)^{20}\mathrm{Ne}$ and $^{23}\mathrm{Na}(\mathrm{p},\gamma)^{24}\mathrm{Mg}$ determine if indeed the NeNa chain can have a cycling character. Figure 8 indicates that the former reaction is predicted to be always quicker than the latter one, which ensures that the NeNa chain is in fact a cycle. The same conclusion is reached with the use of the CF88 rates.

# 4 The MgAl Chain

The MgAl chain is illustrated in Fig. 5. It involves in particular $^{26}$Al. Its long-lived ($t_{1/2} = 7.05\times10^5$ y) $^{26}\mathrm{Al}^{\mathrm{g}}$ ground state and its short-lived ($t_{1/2} = 6.35$ s) $^{26}\mathrm{Al}^{\mathrm{m}}$ isomeric state have to be considered as two separate species at the temperatures of relevance for the non-explosive H burning (Ward and Fowler, 1980).

The rates of several reactions involved in the MgAl chain are now put on firm grounds (see e.g. Arnould and Mowlavi 1993 for references). In spite of much recent effort (Vogelaar 1989, Champagne et al. 1993, Vogelaar et al. 1995), one noticeable exception concerns $^{26}\mathrm{Al}^{\mathrm{g}}(\mathrm{p},\gamma)^{27}\mathrm{Si}$. The remaining large uncertainties in the considered temperature range are illustrated in Fig. 9. One notes that the uncertainties re-evaluated recently by one of us (A.C.), and leading to the rate referred to as CH95 in Fig. 9, are much larger than the CH93 ones. It has to be emphasized that the newly estimated boundaries rely on theoretical arguments only. Although this evaluation is thought to be reliable, the measured rate might ultimately be found to lie outside of this suggested range[1]. In the other hand, Fig. 10 illustrates the uncertainties predicted by the CH95 new re-evaluation of the $^{26}\mathrm{Mg}(\mathrm{p},\gamma)^{27}\mathrm{Al}$ rate.

The yield predictions for the nuclides involved in the MgAl chain are presented in Fig. 11. It is seen in particular that $^{25}$Mg can be significantly destroyed at H exhaustion for $T_6 \gtrsim 40$, so that $^{26}\mathrm{Al}^{\mathrm{g}}$ can start to be built up substantially in this temperature range. Most unfortunately, its predicted yields get highly unreliable at these very same temperatures, where $^{26}\mathrm{Al}^{\mathrm{g}}(\mathrm{p},\gamma)^{27}\mathrm{Al}$ could start becoming more rapid than the $^{26}\mathrm{Al}^{\mathrm{g}}$ $\beta$-decay (Fig. 9). In such conditions, the proton capture rate uncertainties fully translate into abundance uncertainties. In contrast, at $T_6 \lesssim 40$, the $\beta$-decay dominates, thus obliterating the proton capture rate uncertainties. It is also faster than the H-burning timescale, so that $^{26}\mathrm{Al}^{\mathrm{g}}$ has time to be transformed into $^{26}$Mg by the end of H burning.

Figure 11 indicates that the $^{27}$Al abundances are largely uncertain as well. This relates directly to the above-mentioned uncertainties in the rate of the $^{26}\mathrm{Al}^{\mathrm{g}}(\mathrm{p},\gamma)^{27}\mathrm{Si}$ reaction involved in the production channel $^{26}\mathrm{Al}^{\mathrm{g}}(\mathrm{p},\gamma)^{27}\mathrm{Si}(\beta^+)^{27}\mathrm{Al}$, as the well to the uncertainties still remaining in the CH95 evaluation of the other $^{27}$Al production mode $^{26}$Mg (p, $\gamma$) $^{27}$Al (Fig. 10). On top of the uncertain production of $^{27}$Al, the efficiency of its destruction cannot be reliably predicted either. This comes from the uncertainties remaining in the $^{27}$Al (p, $\alpha$) $^{24}$Mg and $^{27}$Al (p, $\gamma$) $^{28}$Si rates, as displayed in Fig. 12.

[1]The recent analysis of the $^{26}\mathrm{Al}^{\mathrm{g}}(\mathrm{p},\gamma)^{27}\mathrm{Si}$ rate by Vogelaar et al. (1995) leads to a slight reduction of the CH95 upper bound displayed in Fig. 9. More specifically, the effect amounts to less than 10% for $T_6 \leq 50$ and to a factor of about 2 in the $50 < T_6 \leq 80$ range. This reduction does not affect significantly the conclusions drawn here on grounds of the CH95 rates.

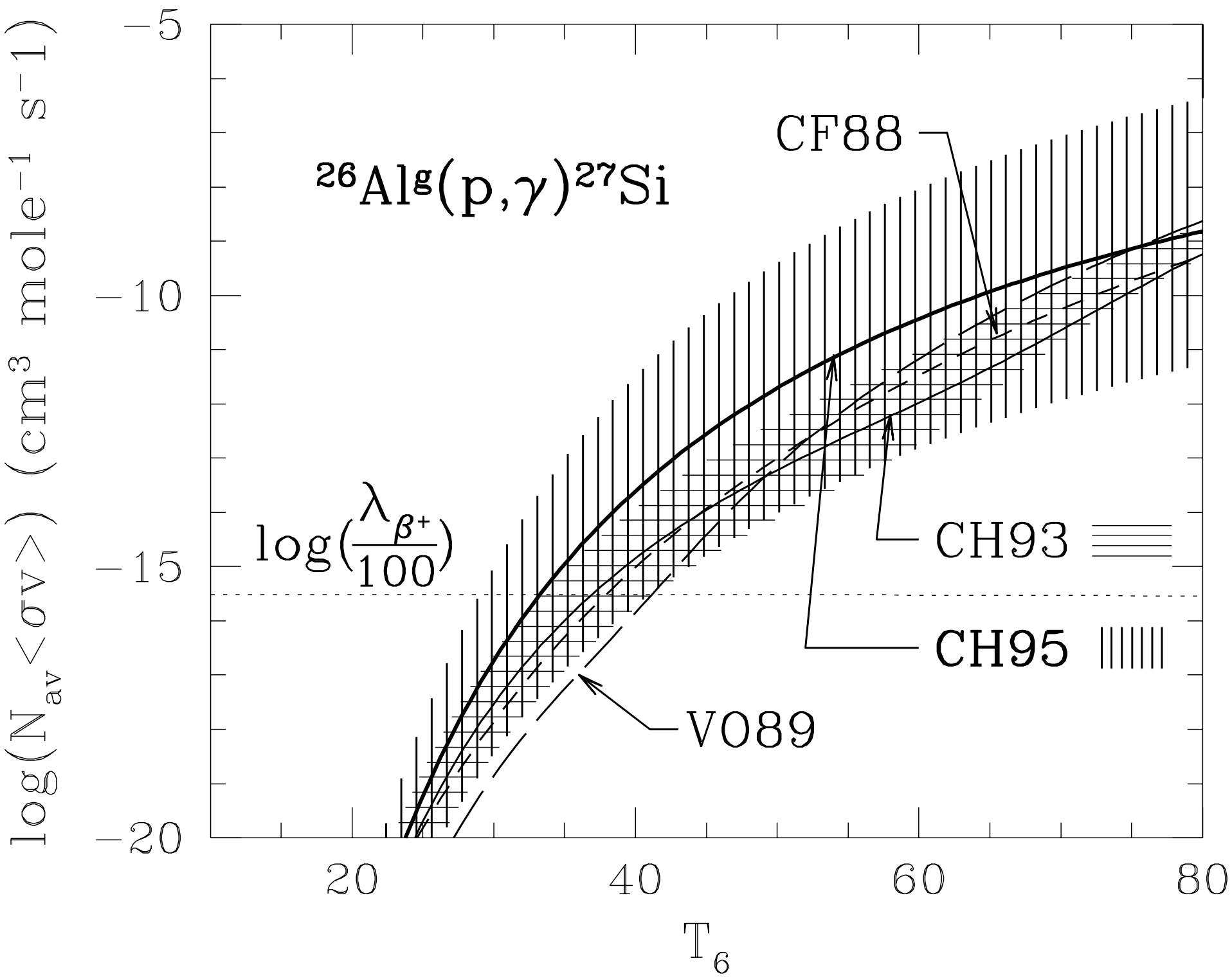

Figure 9: Same as Fig. 3, but for $^{26}$Al$^{g}$(p, $\gamma$)$^{27}$Si. VO89, CH93 and CH95 refer to Vogelaar (1989), to Champagne et al. (1993), and to a recent re-evaluation of the rate conducted by one of us (A.C.). For comparison, the horizontal dotted line indicates the value of $\log(\lambda_\beta/\rho X)$ with $\rho X = 100$ g cm$^{-3}$ ($\lambda_\beta$ is the $^{26}$Al$^{g}$ $\beta$-decay rate, $\rho$ the density and $X$ the H mass fraction).

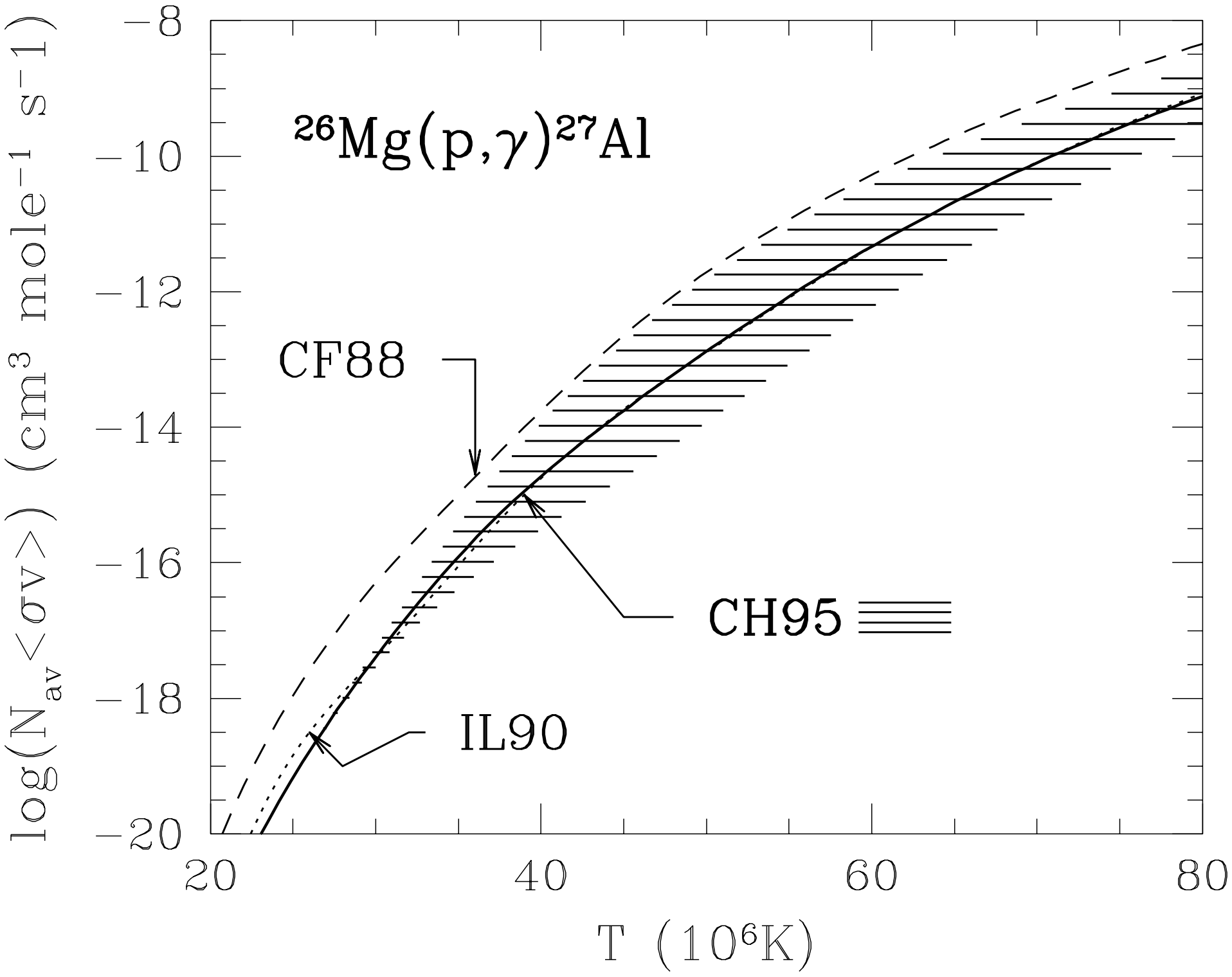

Figure 10: Same as Fig. 3, but for $^{26}$Mg(p, $\gamma$)$^{27}$Al. IL90 and CH95 refer to the rates derived by Iliadis et al. (1990) and from a recent re-analysis conducted by one of us (A.C.).

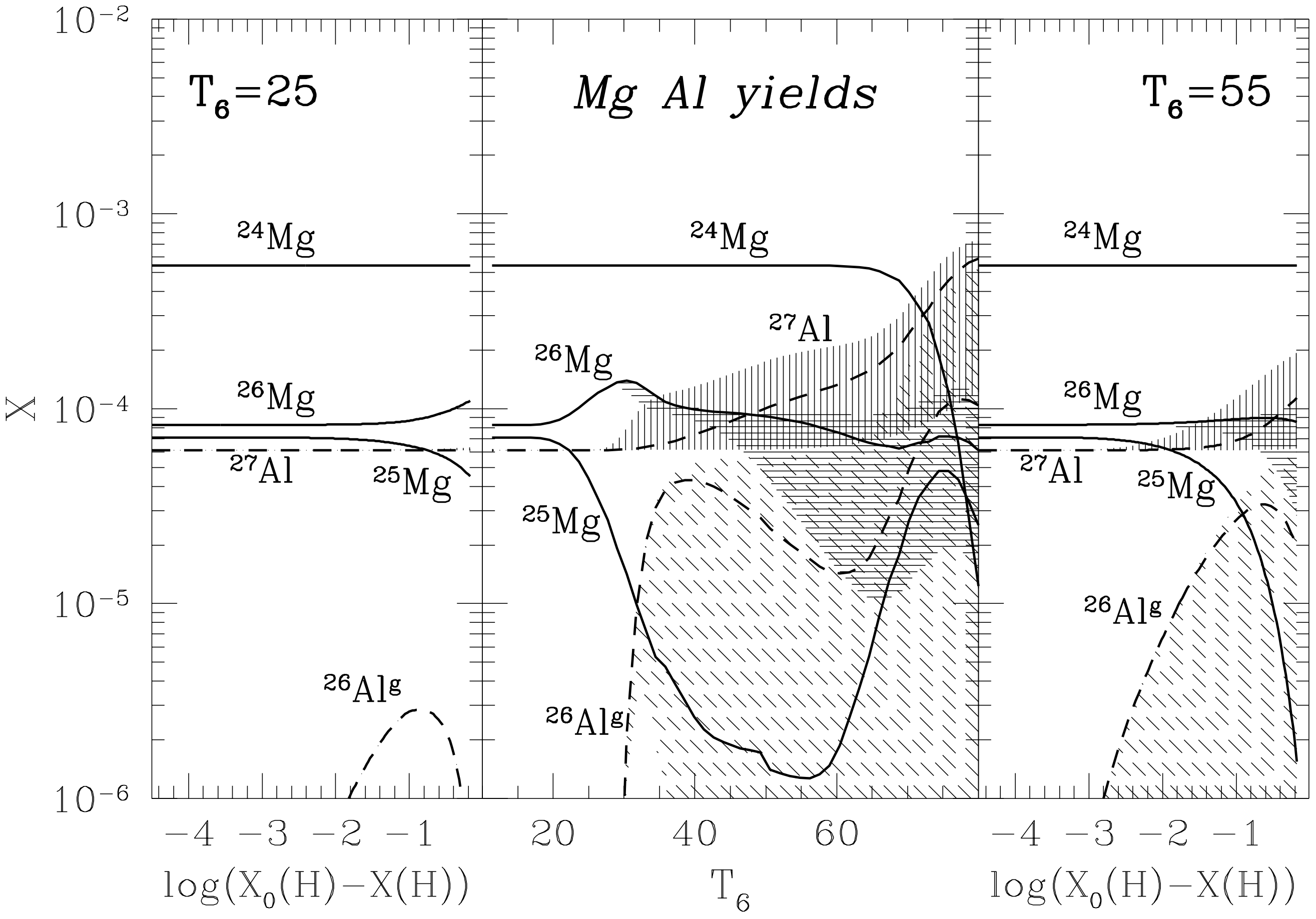

Figure 11: Same as Fig. 2, but for the nuclides involved in the MgAl chain.

The uncertainties in the $^{27}$Al proton capture rates also impact on the prediction concerning the cycling character of the MgAl chain, as determined by the ratio of the $^{27}$Al (p, $\gamma$) $^{28}$Si and $^{27}$Al (p, $\alpha$) $^{24}$Mg rates. From the data displayed in Fig. 13, and when the nuclear uncertainties are duly taken into accout, it appears that a MgAl cycle could possibly set in for $T_6 \lesssim 65$. Clearly, further work is required in order to improve our knowledge of the $^{27}$Al proton capture rates, and thus to specify the extent of the leakage out of the MgAl region.

The large uncertainties in the $^{26}$Al and $^{27}$Al yields that relate to the destruction or production of $^{26}$Al and $^{27}$Al, as well as to the level of cycling of the MgAl chain, are especially unfortunate in view of the prime importance of these two nuclides in cosmochemistry and $\gamma$-ray astronomy. On the one hand, there is now ample observational evidence that $^{26}$Al has decayed in situ in various meteoritic inclusions, where the $(^{26}\mathrm{Al}^g/^{27}\mathrm{Al})_0$ ratio at the beginning of the condensation of the solar system solids has the "canonical" value of about 5 $10^{-5}$ (e.g. Wasserburg 1985). There is also strong observational evidence for its decay in identified single grains of likely stellar origin, where the $^{26}\mathrm{Al}^g/^{27}\mathrm{Al}$ ratio can vary in the quite wide $10^{-5} \lesssim {}^{26}\mathrm{Al}^g/^{27}\mathrm{Al} \lesssim 1$ range (e.g. Anders and Zinner 1993, Nittler et al. 1994). On the other hand, the 1.8 MeV $\gamma$-ray emission observed in the galactic disk is attributed to the decay of about 1.5 $M_\odot$ of $^{26}\mathrm{Al}^g$ that have been present in the interstellar medium over the last $\sim 10^6$ years (e.g. Prantzos and Diehl 1995).

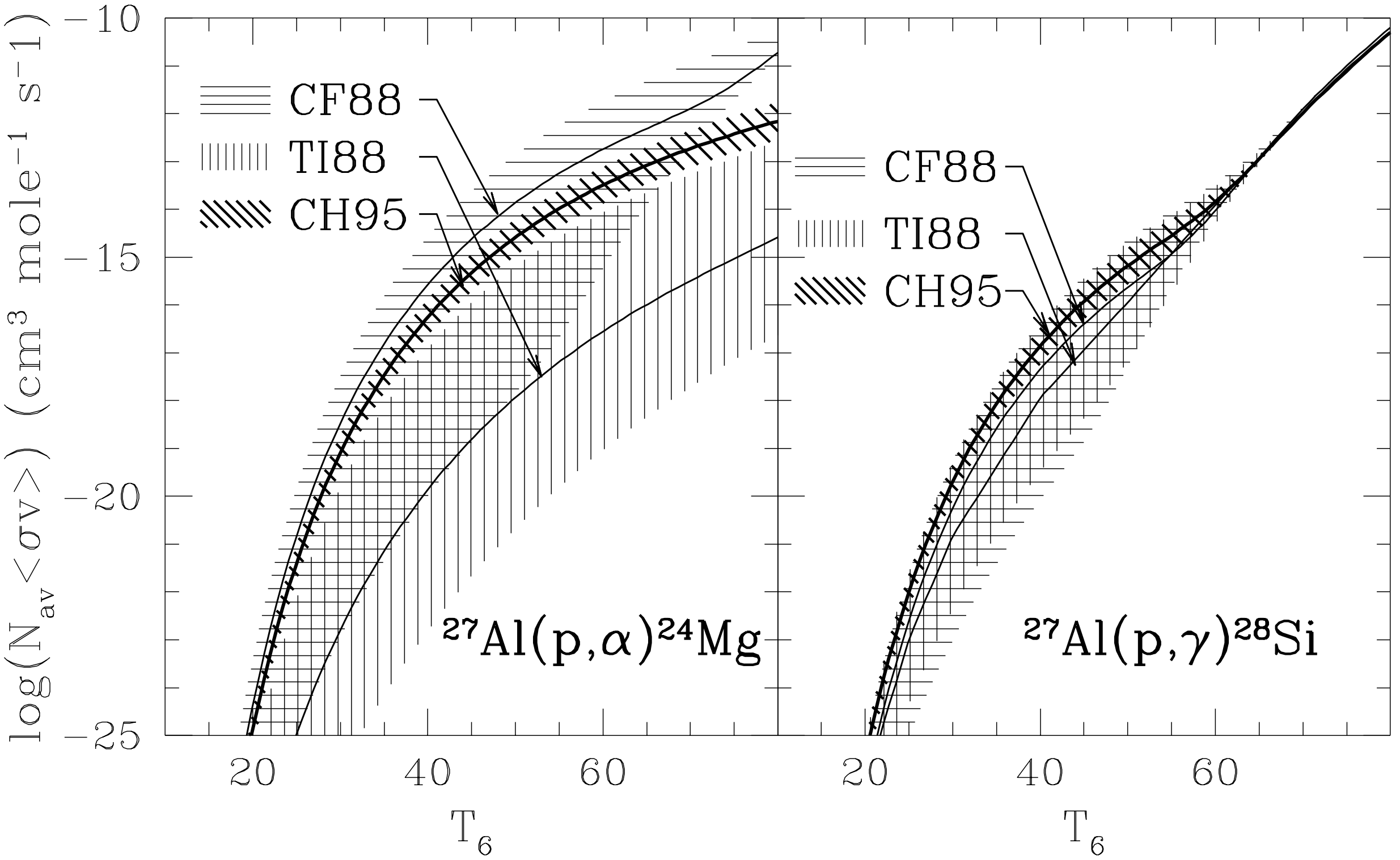

Figure 12: Same as Fig. 3, but for $^{27}Al(p,\alpha)^{24}Mg$ and $^{27}Al(p,\gamma)^{28}Si$. TI88 refers to Timmermann et al. (1988) (see also Champagne et al. 1988), while CH95 refers to a recent re-evaluation conducted by one of us (A.C.).

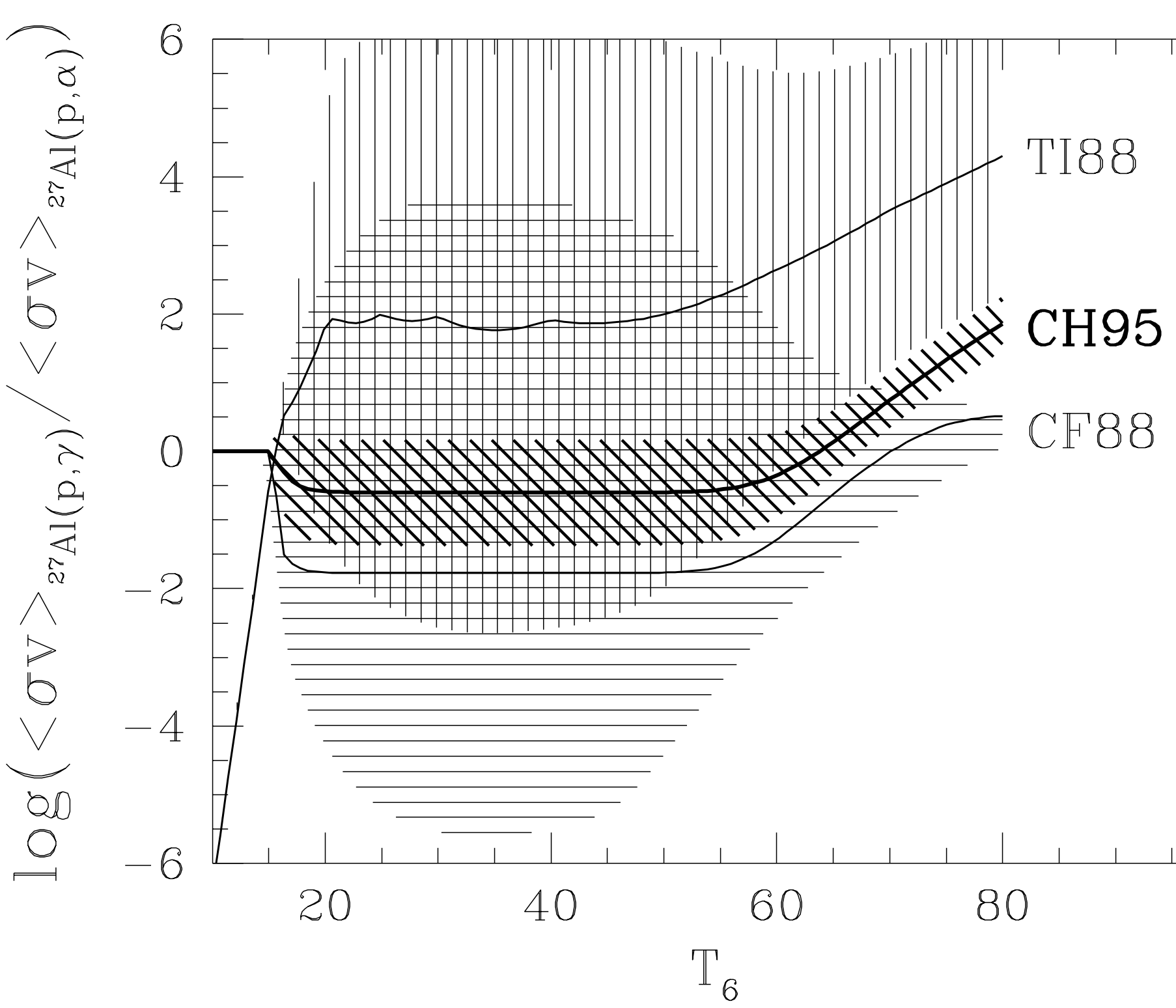

Figure 13: Ratio of the $^{27}Al(p,\gamma)^{28}Si$ to $^{27}Al(p,\alpha)^{24}Mg$ rates. Labels and hatched regions have the same meaning as in Fig. 12.

In fact, Fig. 11 demonstrates that nuclear physics uncertainties prevent any reliable estimate of the $^{26}\mathrm{Al}^g$ yields and of the $^{26}\mathrm{Al}^g/^{27}\mathrm{Al}$ ratios emerging from the MgAl chain if it indeed takes place at temperatures in excess of about $35 \times 10^6$ K. In such conditions, a central question is thus: what are the exact temperatures of operation of the MgAl chain in realistic models for the non-exploding (Asymptotic Red Giant and Wolf-Rayet) stars that have envisioned up to now as possible $^{26}$Al producers? In the case of Wolf-Rayet stars, the relevant MgAl chain operates during core H burning at temperatures that do not exceed about $45 \times 10^6$ K at the very center of the hottest computed stellar cores (Meynet 1994). Peripheral core layers are of course cooler. In such conditions, Fig. 11 suggests that the nuclear uncertainties affecting the predicted Al yields may be of relatively limited extent in a large variety of Wolf-Rayet model stars. In the case of Asymptotic Red Giant stars, $^{26}\mathrm{Al}^g$ is produced either in the thin H burning shell surrounding the electron degenerate core (Forestini et al. 1991), or possibly at the bottom of the convective envelope for the most massive ones. The temperature in the H burning shell can reach 60 to $80 \times 10^6$ K. In view of these rather high temperatures, and considering the results of Fig. 11, the Al yield predictions for Asymptotic Giant stars may thus be put on a less safe nuclear footing than the Wolf-Rayet ones. Of course, astrophysical uncertainties may blur the picture further.

# 5 Conclusions

This brief review makes clear that large uncertainties of nuclear origin still prevent a reliable prediction of the yields of various nuclides of great astrophysical interest involved in the CNO, NeNa and MgAl modes of hydrostatic hydrogen burning. Needless to say, myriads of additional nuclear problems are raised by the following (He, C, O-Ne and Si) non-explosive stellar burning phases.

In many instances, present technologies have been pushed to their ultimate limits, and the measurement of the extraordinarily small cross sections of astrophysical interest requires to bring into operation more imaginative techniques of higher performance than the ones used up to now. Concomitantly, better nuclear models with improved predictive power are urgently called for.

Over the last decades, the dedicated and collaborative work of astrophysicists and experimental or theoretical nuclear physicists has greatly helped improving our understanding of the Universe. No doubt that their future common adventure will be even more exciting and rewarding.